\documentclass[floatfix,aps,prb,amsmath,amssymb,superscriptaddress,notitlepage,twocolumn,reprint]{revtex4-1}
\usepackage{chngcntr}\usepackage{graphicx}
\usepackage{dcolumn}

\usepackage{dirtytalk}

\usepackage{stackengine}
\usepackage[section]{placeins}
\usepackage[dvipsnames*,svgnames]{xcolor}
\usepackage{verbatim}
\usepackage[position=bottom,caption=false,captionskip=3pt,font=normalsize,subrefformat=simple,labelformat=simple]{subfig}

\usepackage{listings}
\usepackage{mathtools}
\usepackage{todonotes}
\usepackage{amsmath}
\usepackage{chngcntr}
\usepackage{bm}

\usepackage{algorithm}
\usepackage{algpseudocode}
\usepackage{url}
\usepackage[super]{nth}

\usepackage{placeins}

\definecolor{GRgreen}{rgb}{0.2, 0.55, 0}
\definecolor{AGpurple}{rgb}{0.5, 0.1, 0.7}
\newcommand{\AG}[1][]{\textcolor{AGpurple}}
\newcommand{\red}[1][]{\textcolor{red}}
\newcommand{\GR}[1][]{\textcolor{GRgreen}}
\newcommand{\AH}[1][]{\textcolor{blue}}
\usepackage{nicefrac}       
\usepackage{multirow}

\usepackage[version=3]{mhchem}
\usepackage{siunitx}
\usepackage{microtype}      
\usepackage{lipsum}
\usepackage[breaklinks]{hyperref}
\usepackage{tabularx}
\usepackage{soul}
\hypersetup{
colorlinks,
linkcolor={blue!100!black},
citecolor={blue},
urlcolor={blue!80!black}}

\newcommand\varpm{\mathbin{\vcenter{\hbox{%
  \oalign{\hfil$\scriptstyle+$\hfil\cr
          \noalign{\kern-.3ex}
          $\scriptscriptstyle({-})$\cr}%
}}}}
\newcommand\varmp{\mathbin{\vcenter{\hbox{%
  \oalign{$\scriptstyle({+})$\cr
          \noalign{\kern-.3ex}
          \hfil$\scriptscriptstyle-$\hfil\cr}%
}}}}

\usepackage{array}

\usepackage{booktabs}

\def\equationautorefname~#1\null{#1\null}

\begin{document}

\title{Universal Effective Medium Theory to Predict the Thermal Conductivity in Nanostructured Materials}

\author{S. Aria Hosseini}

\affiliation{Department of Mechanical Engineering, University of California, Riverside, 900 University Avenue, Riverside, CA 92521, USA}

\author{Sarah Khanniche}

\affiliation{Department of Chemical Engineering, Massachusetts Institute of Technology, 77 Massachusetts Avenue,
Cambridge, Massachusetts 02139, USA}

\author{P. Alex Greaney}
\email{greaney@ucr.edu}

\affiliation{Department of Mechanical Engineering, University of California, Riverside, 900 University Avenue, Riverside, CA 92521, USA}

\author{Giuseppe Romano}
\email{romanog@mit.edu}

\affiliation{Institute for Soldier Nanotechnologies, Massachusetts Institute of Technology, 77 Massachusetts Avenue,
Cambridge, Massachusetts 02139, USA}

\begin{abstract}

Nanostructured materials enable high thermal transport tunability, holding promises for thermal management and heat harvesting applications. Predicting the effect that nanostructuring has on thermal conductivity requires models, such as the Boltzmann transport equation (BTE), that capture the non-diffusive transport of phonons. Although the BTE has been well validated against several key experiments, notably those on nanoporous materials, its applicability is computationally expensive. Several effective model theories have been put forward to estimate the effective thermal conductivity; however, most of them are either based on simple geometries, e.g., thin films, or simplified material descriptions such as the gray-model approximation. To fill this gap, we propose a model that takes into account the whole mean-free-path (MFP) distribution as well as the complexity of the material's boundaries in infinitely thick films with extruded porosity using \emph{uniparameter} logistic regression. We validate our approach, which is called the ``Ballistic Correction Model" (BCM), against full BTE simulations of a selection of three base materials (GaAs, InAs, and Si) with nanoscale porosity, obtaining excellent agreement. While the key parameters of our method, associated with the geometry of the bulk material, are obtained from the BTE, they can be decoupled and used in arbitrary combinations and scales. We tabulated these parameters for a few cases, enabling the exploration of systems that are beyond those considered in this work. Providing a simple yet accurate estimation of thermal transport in nanostructures, our work sets out to accelerate the discovery of materials for thermal-related applications.

\end{abstract}


\maketitle

\section{Introduction}

Engineering semiconducting nanostructures has enabled unprecedented control on nanoscale heat flow for a wide range of applications, from microelectronic devices \cite{martin2006nanotechnology,mateu2005review} to optoelectronics \cite{grundmann2002nano,bharti2013novel} and thermoelectrics (TE) \cite{shakouri2009nanoengineered,shakouri2011recent,snyder2011complex,minnich2009bulk,ouyang2019emerging}. 
Successful control of thermal transport in TE nanomaterials has been shown in thin films, \cite{cheaito2012experimental,braun2016size} superlattices, \cite{hu2012si,mu2015ultra,garg2013minimum} nanowires, \cite{li2012thermal,li2012thermal} nanomeshes, \cite{feng2016ultra,perez2016ultra} nanocomposites, \cite{miura2015crystalline,liao2015nanocomposites}low-dimensional nanomaterials, \cite{zhang2020size, yu2021perspective} and nanoporous structures, \cite{shi2018polycrystalline,PhysRevB.100.035409,PhysRevB.102.205405} all of them featuring extremely low thermal conductivity. 
In multiphase materials that have macro- or micro- scale heterogeneity, heat transfer is often modeled by replacing the detailed microstructure with a homogeneous \emph{effective medium} that has thermal conductivity matching the macroscopic thermal conductivity of the real material. For coarse microstructures, where heat conduction is diffuse and Fourier's law holds, analytical models exist to supply the thermal conductivity of this effective medium based on the volume fraction and conductivity of the constitutive phases in the material it replaces. For example, the effective medium thermal conductivity of materials containing cylindrical pores is given by the Maxwell-Garnett theory, \cite{hasselman1987effective} which predicts the thermal conductivity suppression $(1-\phi)/(1+\phi)$, due to the presence of pores with a total volume fraction $\phi$.

While accurate models exist for the diffusive regime, conceiving their nanoscale counterparts is more challenging. In fact, non-diffusive thermal transport, which is captured by the Boltzmann transport equation (BTE),~\cite{harter2019prediction, harter2020predicting} needs to include the whole distribution of phonon mean-free-paths (MFP)---an aspect that has been difficult to capture in analytical models. For this reason, early approaches are based on single-MFP approximations, e.g., gray material approximation~\cite{prasher2006transverse} or simple geometries~\cite{yang2013mean}.

In this work, we fill this gap by providing a simple analytic model for the effective thermal conductivity, $\kappa_{\mathrm{eff}}$, that includes the full MFP distribution and is suitable from the nano- to the macro- scales and for complex structures. Our approach, referred to as the ``Ballistic Correction Model'' (BCM), is based on the logistic approximation of both the bulk cumulative thermal conductivity and the phonon suppression function. In practice, it is a simple formula that provides $\kappa_{\mathrm{eff}}$ given the relevant geometry's feature size, the characteristic bulk MFP, and the macroscopic suppression factor. All these parameters are computed for a few cases, using the BTE and Fourier's law, and tabulated, enabling a wide space of systems to be explored. We validate the BCM model against BTE calculations on Si, GaAs and InAs, obtaining excellent agreement in most cases. 

The paper is structured as follows. We first report on BTE calculations, applied to Si, GaAs, and InAs membranes with different temperatures and porosities. Then, we detail on the building blocks leading to the BCM model, provided in different sections: i) the derivation of a reduced-order model for $\kappa_{\mathrm{eff}}$, based on descriptors for bulk materials and geometries, and validated agaisnt BTE simulations, ii) the definition of a scale- and material- independent phonon suppression function, and, finally, iii) a procedure to assess $\kappa_{\mathrm{eff}}$ based on precomputed descriptors. The proposed model may help identify novel nanostructures with minimal computational efforts, thus accelerating the search of materials for thermal-related applications.

\section{BTE modeling}

We employ the recently developed anisotropic-MFP-BTE (a-MFP-BTE), implemented in OpenBTE~\cite{romano2021openbte}; this model solves the BTE for uniformly spaced vectorial MFPs, $\mathbf{F}_{ml}$,~\cite{romano2021} i.e.
\begin{equation}\label{bte}
    \mathbf{F}_{ml}\cdot \nabla \Delta T_{ml}^{(n)} + \Delta T_{ml}^{(n)} =\sum_{m'l'} \alpha_{m'l'} \Delta T_{m'l'}^{(n-1)};
\end{equation}
the unknowns $\Delta T_{ml}$ are the pseudo-temperatures of phonons with MFP and direction labeled by $m$ and $l$, respectively. The terms $\mathbf{F}_{ml}=\Lambda_m \mathbf{\hat{s}}_l$ depends on the MFP $\Lambda_m$ and phonon direction $\mathbf{\hat{s}}_l = \sin(\phi_l)\mathbf{\hat{x}}+\cos(\phi_l)\mathbf{\hat{y}}$. The right hand side of Eq.~\ref{bte} can be regarded as the pseudo lattice temperature $\Delta T_L$. Equation~\ref{bte} is solved iteratively, where the first guess is given by the standard Fourier equation. Upon convergence, $\kappa_{\mathrm{eff}}$ is estimated as
\begin{equation}\label{kkeff}
    \kappa_{\mathrm{eff}} = -\frac{L}{\Delta T A}\int \mathbf{J}\cdot \mathbf{\hat{n}} dS = \sum_{ml}\kappa_{ml}S_{ml},
\end{equation}
where $L$ is the periodicity, $A$ is the surface area, $\mathbf{J}=\sum_{ml}\mathbf{G}_{ml} \Delta T_{ml}$ is the heat flux and $S_{ml}$ the suppression function. The terms $\alpha_{ml}$, $\mathbf{G}_{ml}$ and  $\kappa_{ml}$ depend on the intrinsic scattering times and heat capacities, which are computed from first-principles, using \textit{AlmaBTE}. \cite{carrete2017almabte} Specifically, the phonon dispersions as well as the scattering times were computed on a $30 \times 30 \times 30$ point Brillouin zone mesh; the second and third-order interatomic force constants for pristine materials, computed with density functional theory using the virtual crystal approximation, are obtained from the \textit{AlmaBTE} materials database. \cite{carrete2017almabte} The suppression function, $S_{ml}$, along the x-axis is given by
\begin{equation}\label{supp}
   S_{ml} = -\frac{L}{\Delta T A}\int \frac{\Delta T_{ml}-\Delta T_L}{F_{ml,x}} dS =  \frac{\mathbf{\hat{s}}_l\cdot\nabla \Delta \bar{T}_{ml}}{\mathbf{\hat{s}}_{l}\cdot\nabla \Delta T_{\mathrm{ext}}},
\end{equation}
where $\Delta \bar{T}_{ml} = A^{-1}\int \Delta T_{ml} dS$ and the external gradient is $\nabla \Delta T_{\mathrm{ext}} = L^{-1}\Delta T_{\mathrm{ext}} \mathbf{\hat{x}}$. The angularly-averaged suppression function, $S_m = \sum_{l} S_{ml}$, is used throughout the text. Generally, the label $l$ refers to both the polar and azimuthal angles. However, in this work we consider infinite thickness, thus the vectorial MFP discretization is performed in polar coordinates, i.e., the polar angel defined in the XY plane and measured from the y axis.

\begin{figure}[t]
\centering
\includegraphics[width=0.45\textwidth]{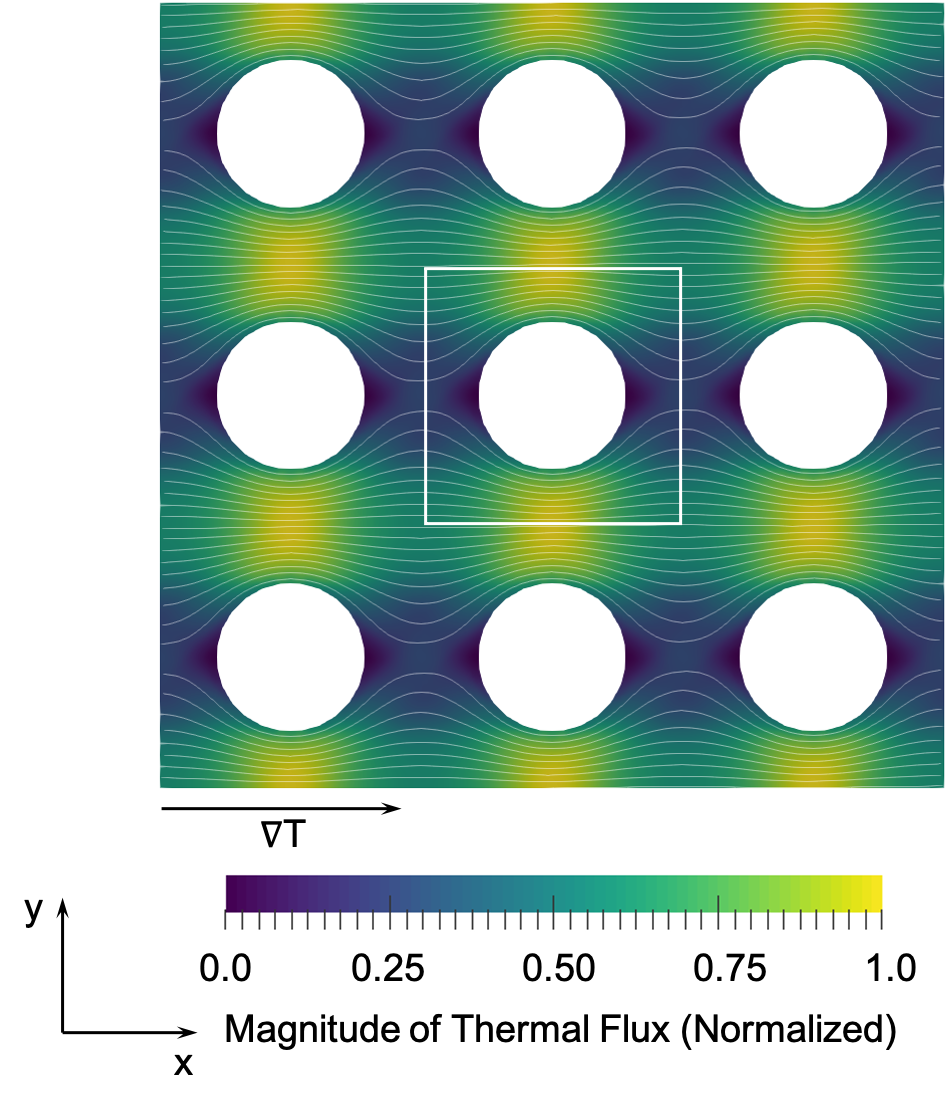}
\caption{Magnitude of the thermal flux in GaAs containing aligned cylindrical pores at an average temperature of 300 K. The system has porosity $\phi=0.25$, the spacing between pores is $\mathrm{L=50\ nm}$ and the white box shows the unit cell. The heat flux is higher in the constriction between the pores, and in addition to creating the constriction, the pores also exert a drag on the heat flux in this region.} 
    \label{fig:1}
\end{figure}

We have computed the effective thermal conductivity, $\kappa_{\mathrm{eff}}$, of InAs, GaAs, and Si membranes containing an array of cylindrical nanopores with different shapes, sizes, and spacings which we parametrize in terms of the total pore fraction, $\phi$, and the array periodicity, $\mathrm{L}$. Figure~\ref{fig:1} shows the distribution of thermal flux for one such case, GaAs with porosity $0.25$ and $\mathrm{L=50\ nm}$.
As we expect, most of the heat is concentrated in the space between the pores. Figures \ref{fig:1} and \ref{fig:2.4} show plots of $\kappa_{\mathrm{eff}}$ of GaAs for a pore spacing of L = 100 nm with porosities of $0.05$ and $0.25$ over a wide range of temperatures. We note that the conductivities obtained from BTE are significantly lower than those from Fourier's law due to the effect of ballistic scattering of long mean free path phonons which is not captured in Fourier's law. The thermal conductivity obtained using Fourier's law only depends on the pore fraction, not the pore spacing, and matches the macroscopic effective medium theory, $\kappa_{\mathrm{eff}}/\kappa_{\mathrm{bulk}} = (1-\phi)/(1+\phi)$. We will use these simulations to validate our proposed model, as detailed in the next sections.

\section{Effective Medium Theory for Nongray Materials}

In this section, we present a reduced-order model for approximating $\kappa_{\mathrm{eff}}$ in nanostructures with aligned pores. Assuming an isotropic intrinsic MFP distribution, $\mathbb{K}(\Lambda)$, and working with continuum variables, the effective thermal conductivity of materials after nanostructuring is written in terms of the suppression of the material's thermal conductivity distribution
\begin{equation}\label{descriptor}
    \kappa_{\mathrm{eff}} = \int_0^{\infty} \mathbb{K}(\Lambda) S(\Lambda)  d\Lambda.
\end{equation}
In Eq.~\ref{descriptor}, $S(\Lambda)$ is the phonon suppression function, a tool describing the degree of reduction of heat transport with respect to the bulk for a given intrinsic MFP $\Lambda$.~\cite{yang2013mean} In the diffusive regime, $S(\Lambda \rightarrow 0) \approx \frac{1-\phi}{1+\phi}$,\cite{romano2017directional} which is in agreement with Maxwell-Garnett prediction, while in the ballistic regime, $S(\Lambda \rightarrow \infty) \propto L_c \Lambda^{-1}$,~\cite{chenbook} where $L_c$ is the mean line-of-sight between phonon scattering events with the nanostructure. Integrating Eq.~\ref{descriptor} by parts, we obtain
\begin{equation}\label{des}
    \kappa_{\mathrm{eff}} = \kappa_{\mathrm{bulk}} \left[ S(\infty) - \int_0^{\infty} \alpha(\Lambda) g(\Lambda) d\Lambda \right],
\end{equation}
where the normalized cumulative thermal conductivity, $\alpha(\Lambda)$, is defined as
\begin{equation}\label{cumulative}
\alpha(\Lambda) = \frac{1}{\kappa_{\mathrm{bulk}}}\int_0^{\Lambda} \mathbb{K}(\lambda) d\lambda,
\end{equation}
and 
\begin{equation}\label{kernel_g}
g(\Lambda) =  \frac{\partial S(\Lambda)}{\partial \Lambda}.
\end{equation}
We note that
\begin{equation}\label{kernel}
S(\infty) =  S(0) + \int_0^{\infty} g(\Lambda) d\Lambda,
\end{equation}
therefore, Eq.~\ref{des} turns into
\begin{equation}\label{descript}
    \kappa_{\mathrm{eff}} = \kappa_{\mathrm{bulk}} \left[ S(0)  + \int_0^{\infty}  g(\Lambda)\left(1 - \alpha(\Lambda) \right) d\Lambda \right].
\end{equation}
We approximate the cumulative thermal conductivity by a logistic function with logarithmic abscissa
\begin{equation}\label{alpha}
    \alpha(\Lambda) = \frac{1}{1+\frac{\Lambda_o}{\Lambda}}.
\end{equation}
Here the entire distribution is described by a single tuning parameter $\Lambda_o$, the characteristic MFP used to fit the logistic function to the cumulative thermal conductivity.~\cite{li2014shengbte} $\Lambda_o$ is the median MFP of the thermal conductivity distribution $\mathbb{K}(\Lambda)$. This very simple function means that the breadth of the distribution cannot be tuned independently from the width of the distribution, and the fitting function contains no asymmetry. However, despite its inflexibility, this simple function does remarkably well at representing the cumulative conductivity of a wide range of materials of different classes, including covalent crystals and compounds such as Si and InP, disordered alloys such Al$_\mathrm{x}$Ga$_\mathrm{{1-x}}$As, and even ionic compounds with a large unit cell such as LiAlO$_{2}$. The data for the cumulative thermal conductivities and the logistic regressions fitted them are publicly available at the GitHub repository~\cite{Hosseini2021}. 
Moreover, what the logistic approximation for $\alpha(\Lambda)$ lacks in flexibility it makes up for in simplicity, enabling us to derive a compact closed-form expression in what follows.

The suppression function can be derived by adding the resistances due to the diffusive and ballistic transport,~\cite{prasher2006transverse} which leads to
\begin{equation}\label{S_model}
    S(\Lambda) = \frac{S(0)}{1+\frac{\Lambda}{L_c}}.
\end{equation}
The term $S(0)$ in Eq. \ref{S_model} is equivalent to the fractional reduction in thermal conductivity predicted by Fourier's law, $S(0) = \kappa_{\mathrm{fourier}}/ \kappa_{\mathrm{bulk}}$.~\cite{romano2017directional} 
\begin{figure}[hbt!]
\centering
\includegraphics[width=0.45\textwidth]{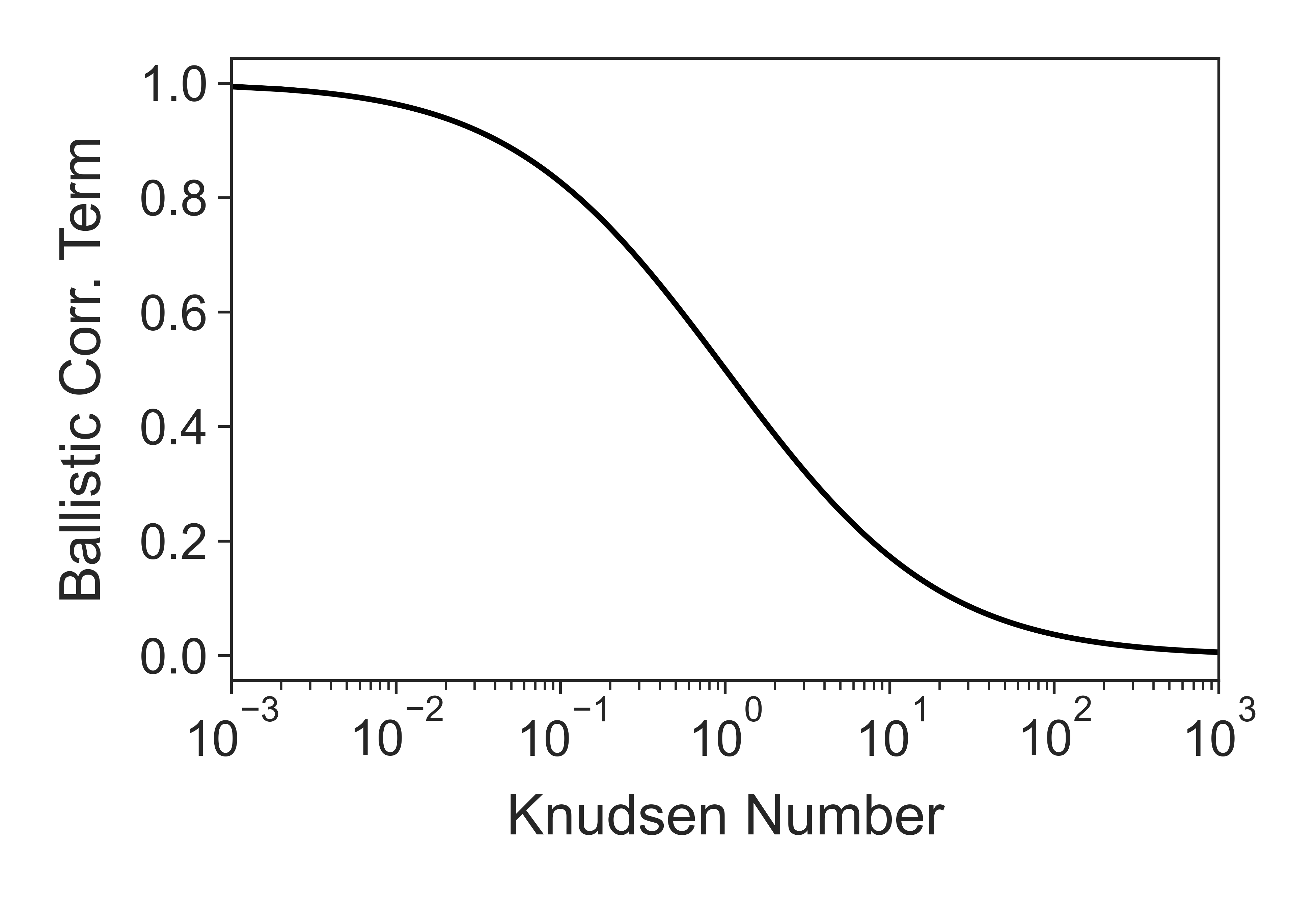}
\caption{Ballistic correction term, $\Xi$ vs. Knudsen number. It can be seen that the scale-dependent correction term becomes significant in the ballistic regime where $Kn \gg 1$.}
    \label{fig:Kn}
\end{figure}
Using Eqs.~\ref{alpha} and \ref{S_model}, Eq.~\ref{descript} becomes
\begin{equation}\label{eq:int}
    \kappa_{\mathrm{eff}}(L_c) = \kappa_{\mathrm{bulk}} S(0) \left[1  - \Lambda_o L_c \int_0^{\infty}  \frac{1}{\left(\Lambda + \Lambda_o\right) \left(\Lambda + L_c\right)^2 } d\Lambda \right],
\end{equation}
which leads to 
\begin{equation}\label{eq:keff}
\kappa_{\mathrm{eff}} = \kappa_{\mathrm{fourier}} \Xi(Kn),
\end{equation}
where the \emph{ballistic correction term} that accounts for truncation of long MFP phonons by nanoscale pores is given by
\begin{equation}\label{eq:Xi}
    \Xi (Kn) = \left[\frac{1 + Kn \left( \ln(Kn) - 1  \right)}{\left ( Kn -1 \right)^2} \right],
\end{equation}
with the Knudsen number $Kn=\nicefrac{\Lambda_o}{L_c}$.

\begin{figure*}[hbt!]

    \centering
    \subfloat[\label{fig:2.1}]{\includegraphics[width=0.5\textwidth]{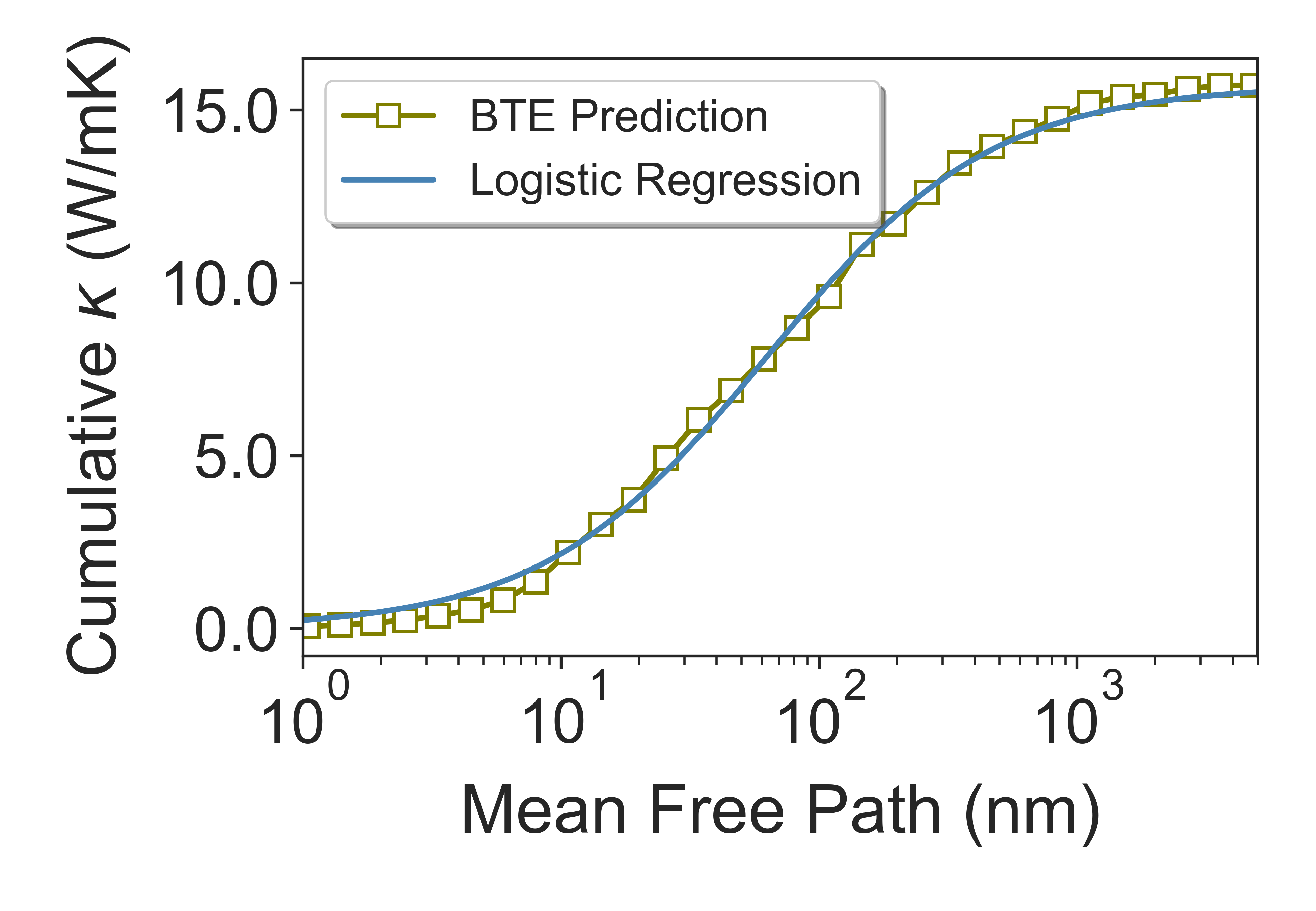}}
    \centering
    \subfloat[\label{fig:2.2}]{\includegraphics[width=0.5\textwidth]{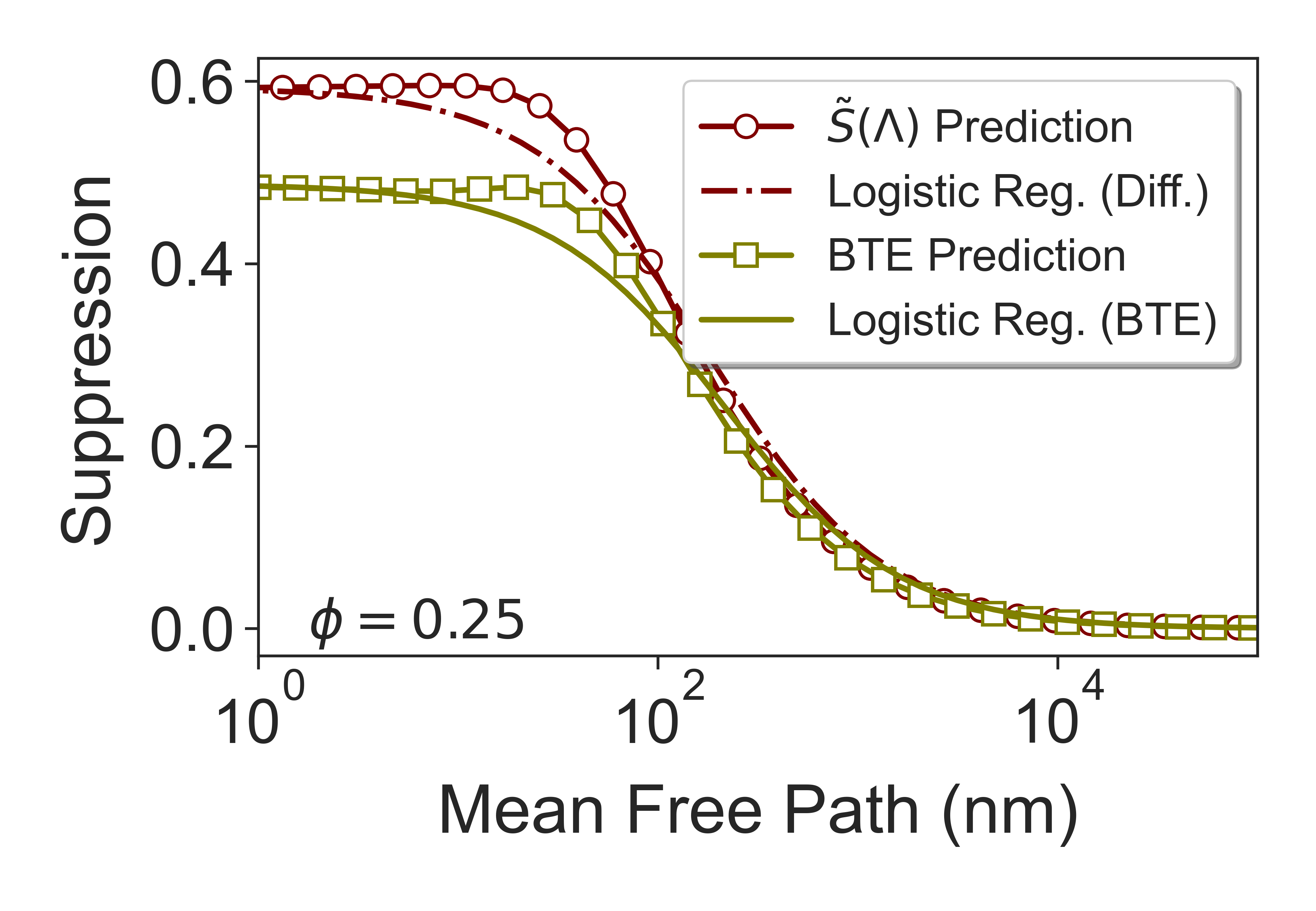}}
        
    \centering
    \subfloat[\label{fig:2.3}]{\includegraphics[width=0.5\textwidth]{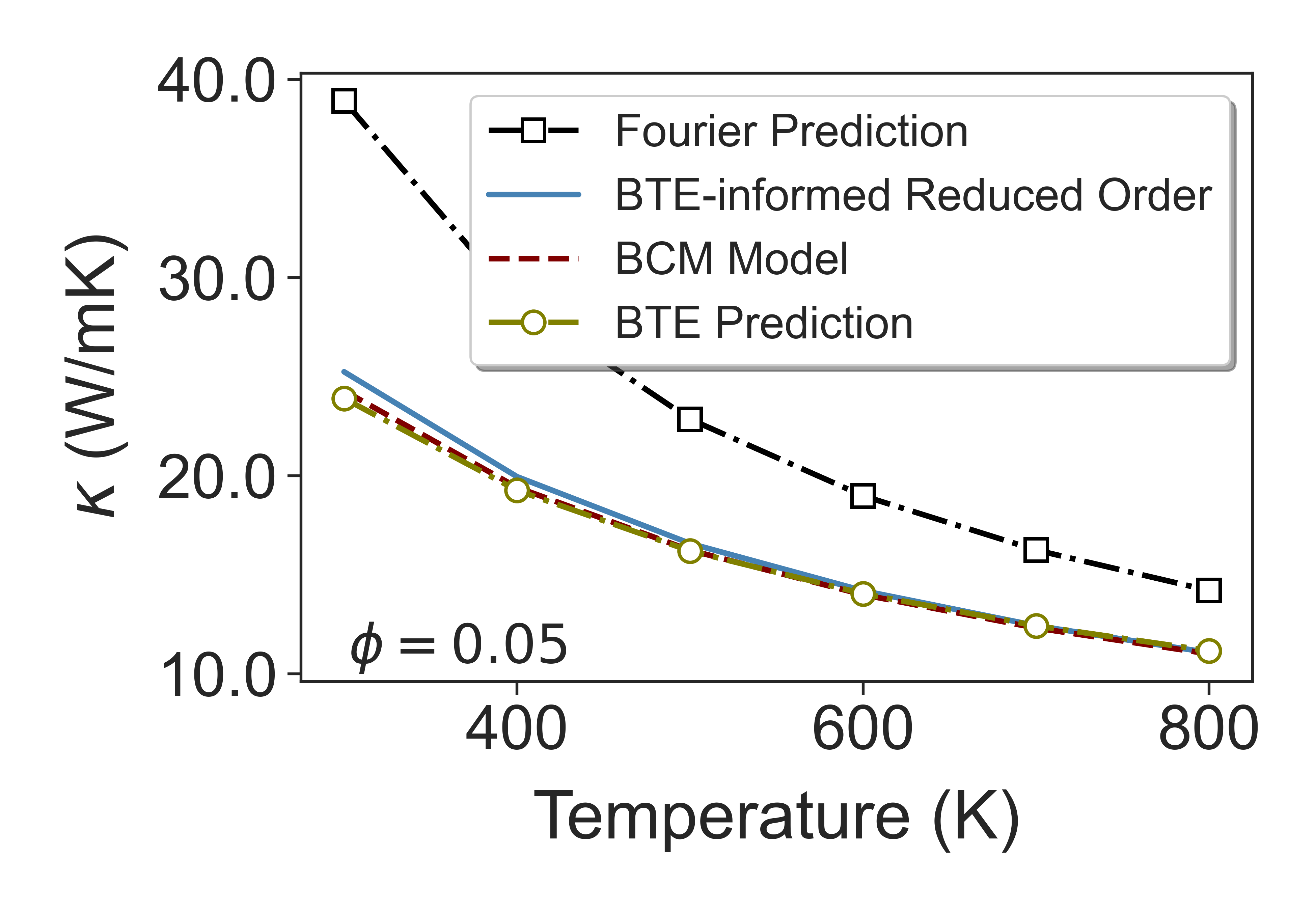}}
    \centering
    \subfloat[\label{fig:2.4}]{\includegraphics[width=0.5\textwidth]{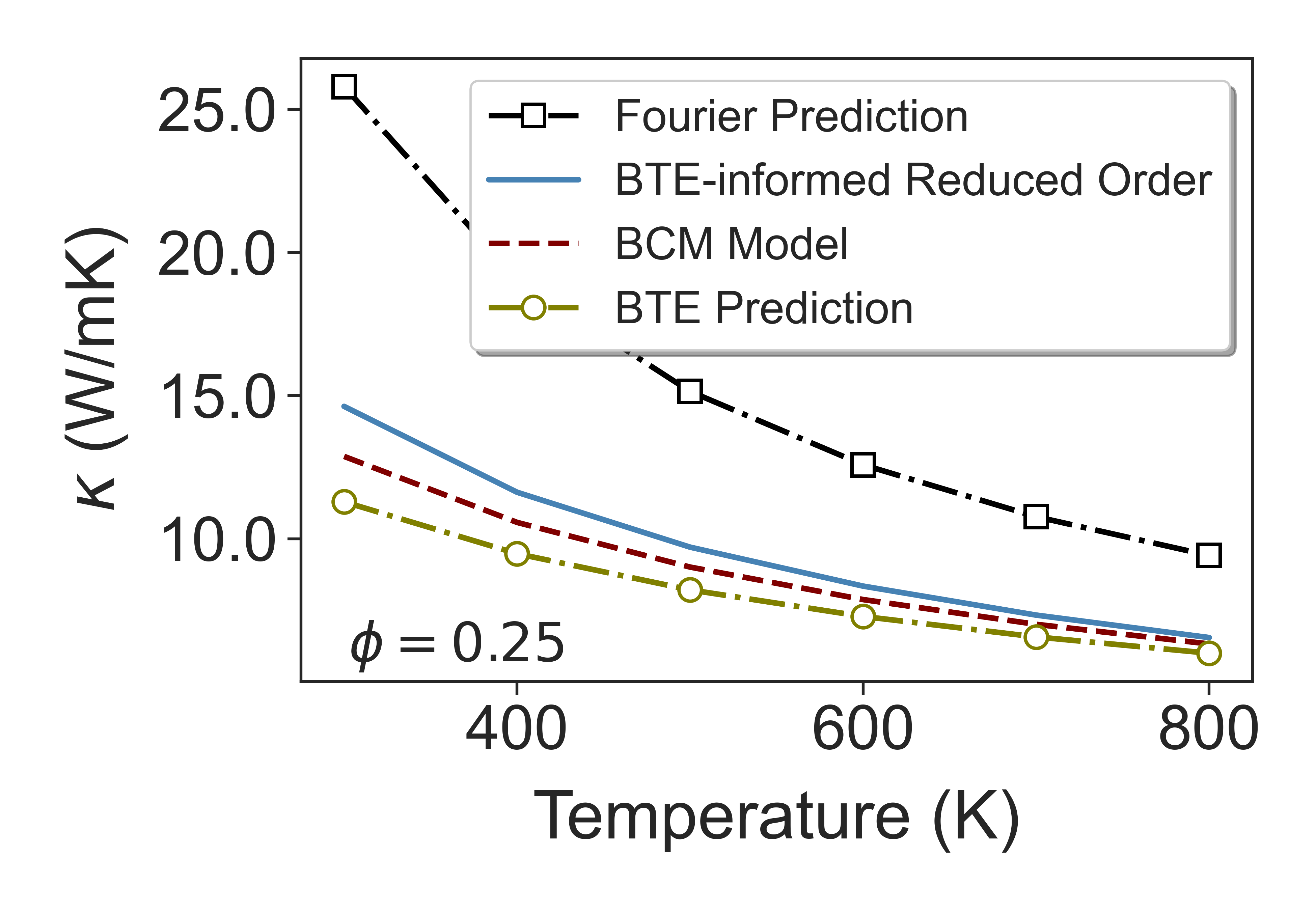}}
    
    \caption{(a) The green squares show the cumulative thermal conductivity of bulk GaAs at 800 K computed using BTE, and the solid blue line shows its least-squares logistic regression fit of Eq. \ref{alpha}. 
    (b) Plot of the suppression function $S(\Lambda,L_c)$ for same GaAs as in (a) but containing an array of cylindrical pores with porosity $\phi = 25\%$ and pore spacing $L= 100$ nm. The plot in green squares shows the suppression function computed from BTE along with its fit of Eq. \ref{S_model} in solid green. The plot in red circles shows the suppression function obtained from the modified BTE from Eq.~\ref{bte-diff} described in section~\ref{sec:diff}, along with its fit of Eq. \ref{S_model} in solid red.
    Panels (c) and (d) plot the thermal conductivity versus temperature for GaAs containing 5\% and 25\% porosity with a 100 nm pore spacing. The black dot-dashed line with square markers shows the prediction from Fourier's law, and the BTE prediction is plotted with a green dot-dashed line with open circles. The prediction from the reduced-order model, Eq~\ref{eq:keff}, using $\tilde{S}$ form of suppression -- see Sec.~\ref{sec:diff} -- is shown in red dash line, and the \emph{BTE-informed} reduced-order model is plotted in solid blue.
    } 
    \label{fig:2}
\end{figure*}
Equations~\ref{eq:keff}--\ref{eq:Xi} are the first main results of our work, and will be referred to as the \emph{BTE-informed} reduced-order model. Interestingly, $\Xi$ depends only on $Kn$. Figure \ref{fig:Kn} shows the ballistic correction $\Xi$ versus Knudsen number. For $Kn = 1$, Eq.~\ref{eq:int} leads to $\kappa_{\mathrm{eff}} = \nicefrac{1}{2}~\kappa_{\mathrm{fourier}}$. For $Kn \rightarrow 0$, $\Xi \rightarrow 1$, recovering the diffusive regime. For large $Kn$, i.e. in the ballistic regime, $\Xi(Kn) \approx \ln{(Kn)} Kn^{-1}$.

The cumulative thermal conductivity of bulk GaAs at room temperature from first principles using \emph{AlmaBTE} and its least-squares logistic regression's fit from Eq.~\ref{alpha} are plotted in figure \ref{fig:2.1}. The logistic curve gives $\Lambda_o = 183\ \mathrm{nm}$. This is roughly the feature size in GaAs-based porous structures where the phonon-pore scattering takes precedence over the anharmonic scattering. 
The suppression function and its regression fit from Eq.~\ref{S_model} are plotted in figure \ref{fig:2.2}. These figures suggest that equations \ref{alpha} and \ref{S_model} are reasonable approximations to $\alpha(\Lambda)$ and $S(\Lambda)$, respectively. 
Figures~\ref{fig:2.3} and~\ref{fig:2.4} illustrate the predictions from the \emph{BTE-informed} reduced-order model applied to the GaAs case for different porosities $\phi$. For $\phi=0.05$, the model agrees with the BTE prediction with less than 5\% error, but with the higher porosity of $\phi=0.25$, the model slightly overestimates the thermal conductivity with about 25\% error at low temperatures and less than 10\% error at high temperatures. The reason for this discrepancy is due to stronger size effects for cases with large porosities, which causes $S(\Lambda)$ to be non-monotonic.~\cite{romano2019diffusive}


\section{The Ballistic Correction Model}\label{sec:diff}

In the previous section, we provide a simple expression for predicting $\kappa_{\mathrm{eff}}$ for different characteristic MFPs, $\Lambda_0$, and material's feature size $L_c$. However, while $\Lambda_0$ can be easily obtained from $\alpha(\Lambda)$ and then used for different geometries, obtaining $L_c$ is more convoluted; in fact, it depends on geometry, periodicity and material. The utility of the reduced-order model for $\kappa_{\mathrm{eff}}$ is only achieved if we have the scheme to estimate the value of $L_c$ efficiently, and thus in this section we set out to achieve this goal. We begin by considering a macroscopic material, for which we have the following approximation
\begin{equation}
    \sum_{m'l'}\alpha_{m'l'} \Delta T_{m'l'}\approx \langle \Delta T_{0l'} \rangle = \Delta T_F,
\end{equation}
where $\langle . \rangle$ denotes an angular average, and $\Delta T_F$ is given by Fourier's law. Equation.~\ref{bte} then becomes
\begin{equation}\label{bte-diff}
      \Lambda_m \mathbf{\hat{s}}_l\cdot \nabla \Delta T_{ml} + \Delta T_{ml} = T_F;
\end{equation}
In this regime, phonon temperatures are isotropic and are given by $\Delta T_{ml}\approx \Delta T_F - \Lambda_m \mathbf{\hat{s}}_l\cdot \nabla \Delta T_F$. The effective thermal conductivity is dominated by the small-MFP limit of the suppression function, e.g.
\begin{equation}
    \kappa_{\mathrm{eff}}\approx \sum_{ml} \kappa_{ml} S_{0l}, 
\end{equation}
with 
\begin{equation}\label{S0}
     S_{0l} = \frac{\mathbf{\hat{s}}_l\cdot\nabla \Delta T_F }{\mathbf{\hat{s}}_{l}\cdot\nabla \Delta T_{\mathrm{ext}}} = \frac{L}{\Delta T_{\mathrm{ext}}} \nabla\bar{T}_F\cdot\left(\hat{\mathbf{x}} + \cot{\phi_l} \mathbf{\hat{y}}\right).
\end{equation}
Using $\sum_l \kappa_{ml}\cot{\phi_l} = 0$, we recover, as expected, the Fourier’s limit, i.e.
\begin{equation}
\kappa_{\mathrm{eff}} \approx  \frac{L}{\Delta T_{\mathrm{ext}}}\nabla T_F \cdot \mathbf{\hat{x}} \sum_{ml}\kappa_{ml}= \kappa_{\mathrm{fourier}}
\end{equation}
In practice, to compute thermal transport in the macroscopic regime, only $\langle S_{0l} \rangle = \kappa_{\mathrm{fourier}}/\kappa_{\mathrm{bulk}}$ is needed, where we used $\int_0^{2\pi}\cot{x}dx = 0$. This quantity is accessible by popular finite-element solvers. Although this derivation took us back to a well-known model for macroscopic heat transport reduction, the fact that it was based on the BTE allows us to define a novel kind of suppression function, $\tilde{S}$, directly computed by the modified BTE from Eq.~\ref{bte-diff}. This quantity was introduced in~\cite{romano2019diffusive} for isotropic MFP distributions. Thus, the above derivation extended it to arbitrary materials, although the main properties remain the same:




i)\ its small-MFP limit corresponds to the suppression predicted by Fourier's law, ii)\ it is material independent, iii)\ it is scale-independent, e.g. it can be translated to consider different dimensions, and vi)\ it always overestimates the suppression functions computed more accurately Eq.~\ref{bte} (i.e., $\tilde{S}(\Lambda)>S(\Lambda)$).
In practice, in our simulation we obtain $\tilde{S}(\Lambda)$ by making the simulation domain large enough, with periodicity being $L_{\mathrm{large}}$, so that $S(0)$ is the same as $\kappa_{\mathrm{fourier}}/\kappa_{\mathrm{BTE}}$, meaning that there are little ballistic effects on small-MFP phonons. Then, we normalize $\Lambda$ by $L_{\mathrm{large}}$, e.g. $\tilde{S}(\Lambda)\rightarrow S(\xi)$, where $\xi = \Lambda L_{\mathrm{large}}^{-1}$. Once $S(\xi)$ is computed for a given geometry, described in terms of relative distances, we assume that the suppression function of a system described by the same geometry and periodicity $L$ is simply given by $S(\Lambda) =\tilde{S}(\Lambda/L)$. Essentially, we are neglecting the effect of non-diffusive phonons on small-MFP phonons even at small scales. Figure~\ref{fig:2.2} shows $\tilde{S}(\Lambda/L)$ with $L$ = 50 nm and porosity $\phi=0.25$. As expected, it deviates from the BTE around the small-MFP region.



Similar to the previous section, we fit $\tilde{S}$ with a logistic function, as shown in \ref{fig:2.2}. In our case, the resulting $L_c$ is 182 nm while that obtained from Eq.~\ref{bte} is 214 nm. As shown from Figs.~\ref{fig:2.3}-\ref{fig:2.4}, the effective thermal conductivities are in good agreement with those computed directly with the BTE. For the low porosity case, the model accurately predicts thermal conductivity (less than 1\% error) but for the high porosity case, the model slightly overestimates the thermal conductivity (less than 10\% error at low temperatures and less than 5\% error at high temperatures). The reason for this trend is that for large size effects, as for the cases with relatively low temperatures or high-porosity, $S$ departs from the logistic function because of the influence of ballistic phonons on small-MFP phonons.~\cite{romano2019diffusive} This effect results in a decrease in $S(0)$ with respect to the Fourier prediction, introducing non-monotonicity in the suppression function. 

\begin{figure}[hbt!]
\centering
\includegraphics[width=0.45\textwidth]{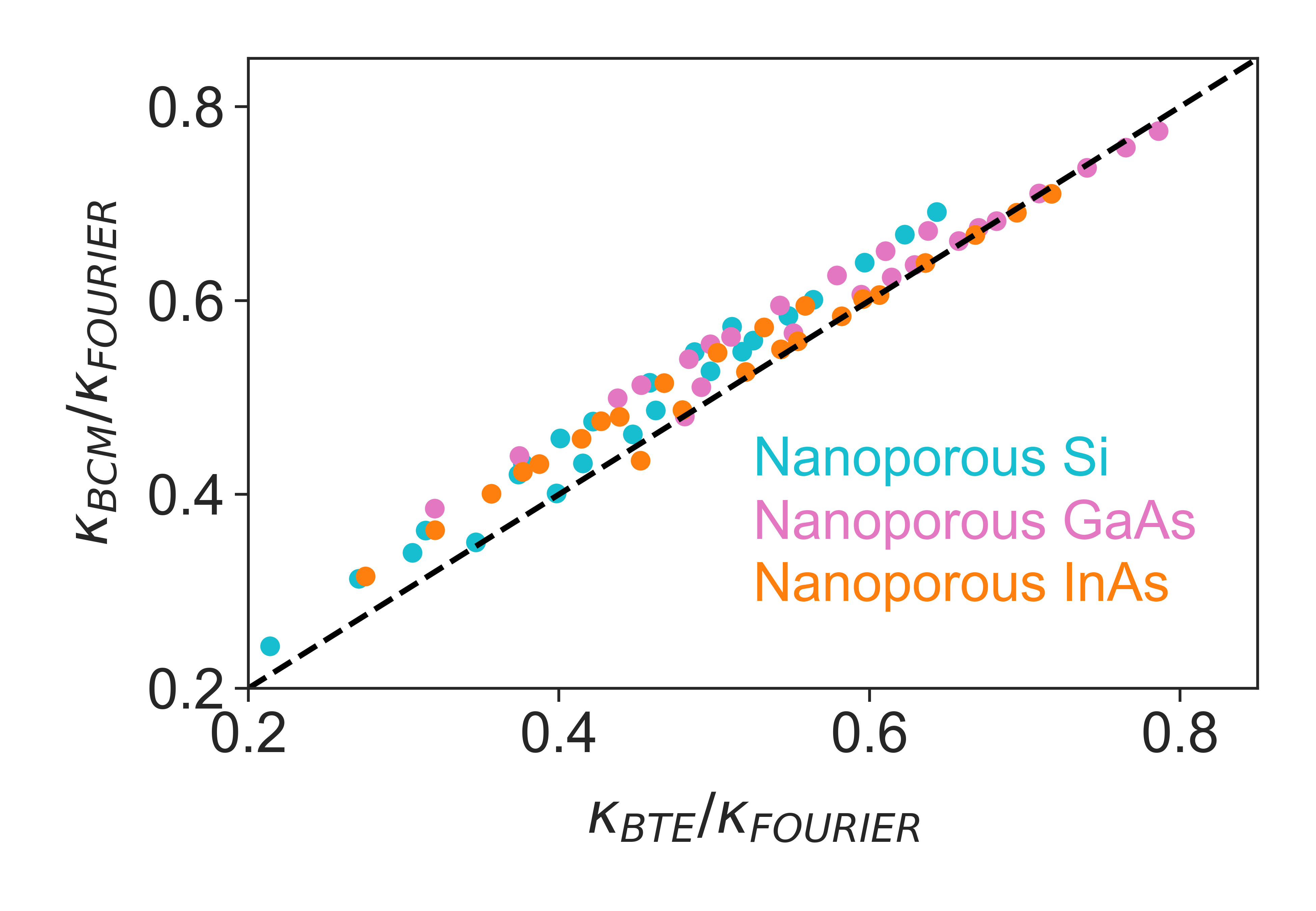}
\caption{Comparison of the thermal conductivity reduction predicted by the reduced-order model  \emph{vs}. the prediction from the full BTE simulation for a selection of three base materials (Si, GaAs, and InAs) at a variety of different temperatures, and containing nanoscale pores with a variety of different porosities and spacings. The periodicity varies from 25--100 nm and the temperature varies from 300--800 K.
}
    \label{fig:results}
\end{figure}

We emphasize that $\tilde{S}$ is material- and scale-independent, hence once $L_c/L$ and $S(0)$ are identified for a given geometry, they can be used for any material and scale, free from tedious Boltzmann transport simulations. To this end, we have pre-computed the key parameters of several bulk materials and geometries in Appendix~\ref{tabulated}. The effective thermal conductivity, once the desired combination is available, can be estimated via the following procedure:
\begin{enumerate}
     \item Select a material at a given temperature (from tables \ref{tbl2} and \ref{tbl3}). Record the associated $\Lambda_o$ and $\kappa_{\mathrm{bulk}}$.
     \item Select a geometry from table \ref{tbl1}. Record associated $S(0)$ and $L_c/L$.
     \item Choose the periodicity of the material, $L$, and compute the associated $L_c$.
     \item Compute the Knudsen number as $Kn = \frac{\Lambda_o}{L_c}$
     \item Compute the effective thermal conductivity as $\kappa_{\mathrm{eff}} = \kappa_{\mathrm{bulk}} S(0) \Xi(Kn)$
\end{enumerate}
This approach denoted as the ``Ballistic Correction Model'' (BCM), is the main result of this paper. We note that, because of the properties of $\tilde{S}$, the BCM provides an upper bound (unless the logistic fit behaves poorly) to $\kappa_{\mathrm{eff}}$, as illustrated in Fig.~\ref{fig:results}.
\section{Conclusion}

To summarize, we have developed a general model to predict lattice thermal conductivity of dielectrics with nanoscale to macroscale porosity. In this model, the cumulative lattice thermal conductivity is approximated by a logistic function regression with a single tuning parameter, $\Lambda_o$ the median phonon mean free path. 
This is a reasonable approximation for a wide range of materials including AlAs, AlN, GaAs, GaN, GaP, InAs, InP, Si, Ge, LiAlO$_2$, and Sn. We have computed the cumulative thermal conductivities at different temperatures from the first principle using \emph{AlmaBTE}. These calculations are publicly available at the GitHub repository~\cite{Hosseini2021}. The effect of scattering at pore interfaces is described by the phonon suppression function. This suppression function can be approximated well with another logistic function curve with two fitting parameters of $S(0)$ and $L_c$. The former parameter describes phonon suppression in the diffusive regime and the latter one describes the characteristic scattering distance imposed on the phonons by porosity. These parameters are tabulated in Table~\ref{tbl1} for pores with different shapes and porosities. The model is robust in providing a good approximating of the results from Boltzmann transport simulations of lattice thermal conductivity for a wide range of pores shapes, sizes, and spacings that span both the diffusive and ballistic regime in infinitely thick films with extruded porosity. This provides a simple yet accurate estimation of thermal transport in nanostructures that can be used to rapidly screen or design materials for a particular thermal task. As such this work provides an important tool to facilitate the design and discovery of materials for thermal-related applications, without explicitly solving the BTE.

\section{Data Availability}
The data that support the findings of this study are available from the corresponding author upon reasonable request.
\section{Conflict of Interest}
The authors declare no conflict of interest

\section{Acknowledgment}

AH and AG acknowledge the financial support of Pacific Northwest National Laboratory
(PNNL). PNNL is operated by Battelle Memorial Institute for the U.S. Department of
Energy under Contract No. DE-AC05-76RL01830. This work was supported by the
National Nuclear Security Administration, Tritium Sustainment Program through contract
with the University of California-Riverside.

\appendix

\section{Tabulated Data for the BCM}\label{tabulated}

\begin{figure*}[hbt!]
\centering
\includegraphics[width=0.8\textwidth]{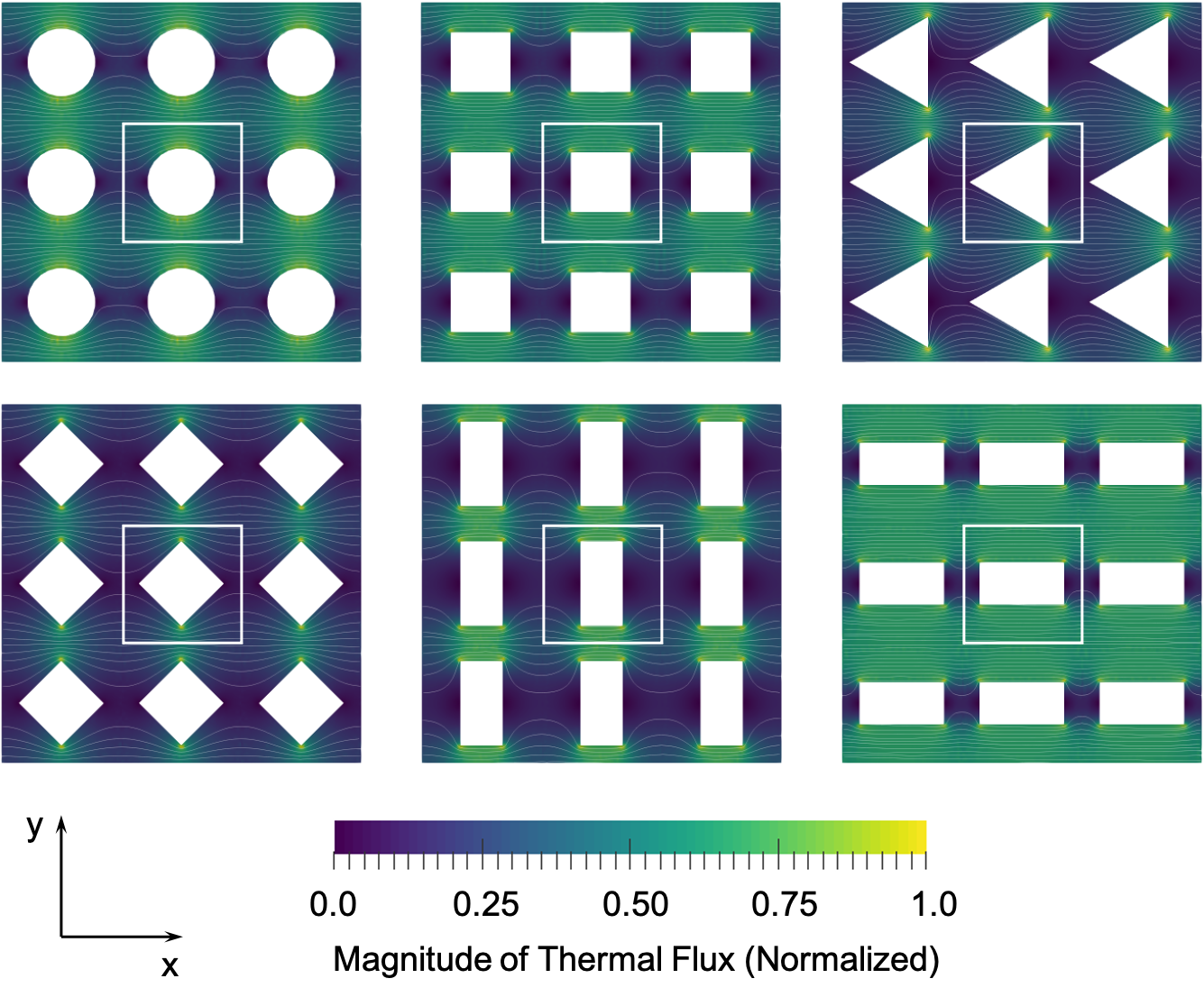}
\caption{Magnitude of the thermal flux in diffusive (material-independent) structure containing aligned pores with different shapes. The porosity is fixed to 0.25 and periodicity is $\mathrm{L=50\ nm}$. The unit cells are highlighted.} 
    \label{fig:a1}
\end{figure*}

Table \ref{tbl1} shows the phonon characteristic length, $L_c$ and diffusive suppression function, $S(0)$ for pores with different shapes and porosities. Note that in case of large porosity with small periodicity and thereby small pore-pore spacing, $S(0)$ may deviate from the Fourier; therefore, the logistic regression is not accurate for the short mean free path regime. Thus, we have limited our calculations up to 50\% porosity. The shapes of the pores are shown in Fig. \ref{fig:a1}. In all cases, the temperature gradient is along the abscissa. Tables \ref{tbl2} and \ref{tbl3} show the characteristic MFP and bulk thermal conductivity of wide sweep of IV and III-V dielectrics at different temperatures.


\begin{table*}[htb!]
\rmfamily
  \caption{The ballistic correction model prediction for phonon characteristic length and diffusive suppression ($S(0)$) in pores with different shapes and porosity. In all cases temperature gradient is along abscissa.}
    \makebox[\linewidth]{
  \begin{tabular}{l*{13}{c}}
    \toprule
\multirow{2}{*}{Porosity}   & \multicolumn{2}{c}{Circle pore} & \multicolumn{2}{c}{Square pore} & \multicolumn{2}{c}{Triangle pore} & \multicolumn{2}{c}{Rectangle pore*} & \multicolumn{2}{c}{Rectangle pore**} & \multicolumn{2}{c}{Rhombus pore}\\ \cmidrule(r){2-3} \cmidrule(l){4-5} \cmidrule(l){6-7} \cmidrule(r){8-9} \cmidrule(l){10-11} \cmidrule(l){12-13}
    & {$\mathrm{L_c/L}$} & {$\mathrm{S(0)}$} & {$\mathrm{L_c/L}$} & {$\mathrm{S(0)}$} & {$\mathrm{L_c/L}$} & {$\mathrm{S(0)}$} & {$\mathrm{L_c/L}$} & {$\mathrm{S(0)}$} & {$\mathrm{L_c/L}$} & {$\mathrm{S(0)}$} & {$\mathrm{L_c/L}$} & {$\mathrm{S(0)}$} \\ [0.5ex]
    \midrule
    \centering{0.05}    & 3.92           & 0.90          & 3.22          & 0.90        & 2.74         &  0.90         & 3.26          & 0.92          & 3.00         & 0.85        & 3.48       &  0.90\\
    \centering{0.10}    & 2.64           & 0.83          & 2.38          & 0.80        & 2.04         &  0.77         & 2.34          & 0.84          & 2.08         & 0.73        & 2.54       &  0.80\\
    \centering{0.15}    & 2.24           & 0.74          & 1.86          & 0.73        & 1.56         &  0.69         & 1.80          & 0.80          & 1.68         & 0.62        & 2.22       &  0.70\\
    \centering{0.20}    & 2.02           & 0.66          & 1.54          & 0.66        & 1.66         &  0.54         & 1.62          & 0.72          & 1.46         & 0.51        & 1.74       &  0.64\\
    \centering{0.25}    & 1.82           & 0.59          & 1.58          & 0.56        & 1.14         &  0.51         & 1.48          & 0.67          & 1.32         & 0.41        & 1.60       &  0.56\\
    \centering{0.30}    & 1.54           & 0.54          & 1.36          & 0.51        & 1.04         &  0.42         & 1.34          & 0.63          & 1.10         & 0.33        & 1.36       &  0.49\\
    \centering{0.35}    & 1.46           & 0.48          & 1.18          & 0.46        & 0.90         &  0.33         & 1.16          & 0.59          & 1.02         & 0.24        & 1.22       &  0.42\\
    \centering{0.40}    & 1.40           & 0.42          & 1.08          & 0.41        & 0.76         &  0.24         & 0.98          & 0.57          & 0.88         & 0.17        & 1.04       &  0.36\\
    \centering{0.45}    & 1.26           & 0.37          & 1.02          & 0.36        & --           &  --           & 0.84          & 0.54          & 0.68         & 0.09        & 1.00       &  0.26\\
    \centering{0.50}    & 1.18           & 0.32          & 0.94          & 0.31        & --           &  --           & 0.78          & 0.50          & --           & --          & --         &  --\\
    \bottomrule
      \label{tbl1}
  \end{tabular}
  }
  * Rectangle pore with the length along the abscissa ($\mathrm{l_x = 2l_y}$) \\
  ** Rectangle pore with the length along the ordinate ($\mathrm{l_y = 2l_x}$)
\end{table*}


\begin{table*}[htb!]

  \caption{The characteristic MFP of IV and III-V dielectrics.}
  \begin{tabular}{l*{11}{c}}
    \toprule
\multirow{2}{*}{Temperature(K)}   & \multicolumn{10}{c}{Characteristic Mean Free Path ($\mu m$)}\\ \cmidrule(r){2-11}
    & {AlAs} & {AlN} & {GaAs} & {GaN} & {GaP} & {Ge} & {InAs} & {InP} & {Si} & {Sn} \\ [0.5ex]
    \midrule
    \centering{200}    & 0.41           & 0.43          & 0.31          & 0.46          & 0.35          & 0.53          & 0.47          & 1.35          & 1.33          & 0.26\\
    \centering{300}    & 0.24           & 0.16          & 0.18          & 0.25          & 0.21          & 0.30          & 0.29          & 0.80          & 0.49          & 0.18\\
    \centering{400}    & 0.17           & 0.10          & 0.13          & 0.18          & 0.16          & 0.21          & 0.21          & 0.58          & 0.29          & 0.13\\
    \centering{500}    & 0.13           & 0.07          & 0.10          & 0.15          & 0.13          & 0.16          & 0.17          & 0.45          & 0.21          & 0.11\\
    \centering{600}    & 0.11           & 0.05          & 0.08          & 0.13          & 0.10          & 0.13          & 0.14          & 0.37          & 0.17          & 0.09\\
    \centering{700}    & 0.09           & 0.04          & 0.07          & 0.11          & 0.09          & 0.11          & 0.12          & 0.32          & 0.14          & 0.08\\
    \centering{800}    & 0.08           & 0.04          & 0.06          & 0.10          & 0.08          & 0.10          & 0.10          & 0.28          & 0.12          & 0.07\\
    \bottomrule
      \label{tbl2}
  \end{tabular}
\end{table*}

\begin{table*}[htb!]

  \caption{The bulk thermal conductivity of IV and III-V dielectrics.}
  \makebox[\linewidth]{
  \begin{tabular}{l*{11}{c}}
    \toprule
\multirow{2}{*}{Temperature(K)}   & \multicolumn{10}{c}{Bulk Thermal Conductivity (W/mK)}\\ \cmidrule(r){2-11}
    & {AlAs} & {AlN} & {GaAs} & {GaN} & {GaP} & {Ge} & {InAs} & {InP} & {Si} & {Sn} \\ [0.5ex]
    \midrule
\centering{200}      & 127   & 594    & 71    & 359     & 122   & 90    & 44     & 126    & 287    & 28\\
\centering{300}      & 81    & 302    & 45    & 250     & 82    & 60    & 28     & 81     & 165    & 19\\
\centering{400}      & 60    & 203    & 34    & 197     & 62    & 45    & 21     & 60     & 118    & 14\\
\centering{500}      & 48    & 155    & 27    & 165     & 50    & 37    & 17     & 48     & 93     & 12\\
\centering{600}      & 40    & 126    & 22    & 142     & 42    & 31    & 14     & 40     & 76     & 10\\
\centering{700}      & 34    & 106    & 19    & 126     & 36    & 27    & 12     & 34     & 65     & 8\\
\centering{800}      & 30    & 92     & 17    & 113     & 32    & 24    & 10     & 30     & 57     & 7\\
    \bottomrule
      \label{tbl3}
  \end{tabular}
  }
\end{table*}

\FloatBarrier
\bibliography{references} 

\end{document}